\documentclass[final,5p,times,twocolumn,square,comma,sort&compress]{elsarticle}

\usepackage{amssymb}
\usepackage{lipsum}
\usepackage{xcolor}
\usepackage{graphicx}
\usepackage{dcolumn}
\usepackage{bm}
\usepackage{ifpdf}
\usepackage{epstopdf}
\usepackage{slashed}
\usepackage{amsfonts}
\usepackage{mathrsfs}
\usepackage{amssymb}
\usepackage{multirow}
\usepackage{CJK}
\usepackage{xcolor}
\usepackage{subfigure,dcolumn}
\usepackage[T2A,T1]{fontenc}
\usepackage[russian,english]{babel}
\usepackage{amsmath}
\usepackage{listings}
\usepackage{booktabs}
\usepackage{color}
\usepackage[switch]{lineno} 
\journal{Physics Letters B}

\begin{document}
% \linenumbers 
\begin{frontmatter}

\title{Bayesian uncertainty quantification for synthesizing superheavy elements}
\author[1]{Yueping Fang}
\author[1]{Zepeng Gao}
\author[1]{Yinu Zhang}
\author[1]{Zehong Liao}
\author[1]{Yu Yang}
\author[1,2]{Jun Su}
\author[1,2]{Long Zhu\corref{mycorrespondingauthor}}
\cortext[mycorrespondingauthor]{Corresponding author: zhulong@mail.sysu.edu.cn}

\affiliation[1]{organization={Sino-French Institute of Nuclear Engineering and Technology, Sun Yat-sen University},
            city={Zhuhai},
            postcode={519082}, 
            state={},
            country={China}}
\affiliation[2]{organization={Guangxi Key Laboratory of Nuclear Physics and Nuclear Technology, Guangxi Normal University},
            city={Guilin},
            postcode={541004}, 
            country={China}}

\begin{abstract}
To improve the theoretical prediction power for synthesizing superheavy elements beyond Og, a Bayesian uncertainty quantification method is employed to evaluate the uncertainty of the calculated evaporation residue cross sections (ERCS) for the first time. The key parameters of the dinuclear system model (DNS-sysu), such as the diffusion parameter \textit{a}, the damping factor $E_\mathrm{d}$, and the level-density parameter ratio $a_\mathrm{f}/a_\mathrm{n}$ are systematically constrained by the Bayesian analysis of recent ERCS data. One intriguing behavior is shown that the optimal incident energies (OIE) corresponding to the largest ERCS weakly depend on the fission process. We also find that these parameters are strongly correlated and the uncertainty propagation considering the parameters independently is not reasonable. The 2\({\sigma}\) confidence level of posterior distributions for $a = 0.586_{-0.002}^{+0.002}$ fm, $E_\mathrm{d} = 25.65_{-3.41}^{+3.43}$ MeV, and $a_\mathrm{f}/a_\mathrm{n} = 1.081_{-0.021}^{+0.021}$ are obtained. Furthermore, the confidence levels of the ERCS and OIE for synthesizing Z = 119 via the reactions \({}^{54}\mathrm{Cr}+{}^{243}\mathrm{Am}\), \({}^{50}\mathrm{Ti}+{}^{249}\mathrm{Bk}\), and \({}^{51}\mathrm{V}+{}^{248}\mathrm{Cm}\) are predicted. This work sets the stage for future analyses to explore the OIE and reaction systems for the synthesis of superheavy elements.

\end{abstract}

\end{frontmatter}

The periodic table of elements has been significantly extended by using the heavy-ion fusion reactions in the past decades \cite{ohrstrom2016names,oganessian2015super,RevModPhys.91.011001,morita2012new,RevModPhys.72.733}. In recent years, to open the eighth period, the worldwide efforts have been made for synthesizing the superheavy elements (SHEs) beyond Og. In an attempt to produce the new element Z = 120, Hofmann et al. at GSI investigated the reaction \({}^{54}\mathrm{Cr}+{}^{248}\mathrm{Am}\) \cite{hofmann2016remarks}, while at the same site, Khuyagbaatar et al. investigated reactions \({}^{50}\mathrm{Ti}+{}^{249}\mathrm{Bk}\) and \({}^{50}\mathrm{Ti}+{}^{249}\mathrm{Cf}\) to produce the new elements Z = 119 and 120 \cite{khuyagbaatar2020search}. Additionally, Oganessian et al. at Dubna attempted to synthesize the SHE with Z = 120 using the reaction \({}^{58}\mathrm{Fe}+{}^{244}\mathrm{Pu}\) \cite{oganessian2009attempt}. Another experiment at RIKEN attempted to synthesize nuclei with Z = 119 by using the combination of \({}^{51}\mathrm{V}+{}^{248}\mathrm{Cm}\) \cite{tanaka2022probing}. Unfortunately, no relevant decay chains were observed due to very low evaporation residue cross sections (ERCS) which strongly depends on the incident energy. Therefore, the precise predictions of the incident energy are crucially important for hunting new SHEs \cite{li2023possibility}.\par

Theoretically, the synthesis of SHEs can be divided into three steps \cite{hagino2018hot}. The ERCS is calculated as the summation over all partial waves \textit{J} \cite{adamian1998fusion}:
\begin{equation}
\begin{aligned}
\sigma_{\mathrm{ER}}(E_{\mathrm{c.m.}})=&\frac{\pi\hbar^{2}}{2\mu E_{\mathrm{c.m.}}}\sum_{J}(2J+1)T(E_{\mathrm{c.m.}},J)\\
&P_{\mathrm{CN}}(E_{\mathrm{c.m.}},J)W_{\mathrm{sur}}(E_{\mathrm{c.m.}},J),
\end{aligned}
\end{equation}
where \(T(E_{\mathrm{c.m.}},J)\) is the transmission probability, \(P_{\mathrm{CN}}(E_{\mathrm{c.m.}},J)\) is the fusion probability and \(W_{\mathrm{sur}}(E_{\mathrm{c.m.}},J)\) denotes the survival probability. Due to the complexity of the theoretical process, particularly the presence of delicate ambiguities in the fusion mechanism \cite{lu2016synthesis,loveland2015experimentalist}, the ERCS calculated by the different theoretical models exhibit significant uncertainty and model dependence \cite{wang2012theoretical,kayumov2022analysis,siwek2012predictions,zhu2014production,deng2023examination}. Even within the same theoretical framework, the predicted ERCS can exhibit considerable discrepancies due to several adjustable parameters \cite{PhysRevC.109.014622,PhysRevC.94.034616}. Therefore, to enhance the predictive power, it is desirable to elucidate the correlations among the key parameters, constrain their ranges, and analyze the associated theoretical uncertainties.\par

Recently, machine learning has been extensively and successfully used in nuclear physics \cite{he2023machine,he2023high,boehnlein2022colloquium,wang2023machine,ma2023phase}. Bayesian inference is a methodology for parameter estimation and uncertainty quantification in a wide range of models used in various fields \cite{huth2022constraining,qiu2024bayesian,li2024importance,xu2021bayesian,wang2019bayesian}. It can be used to constrain physical quantity (or vector of parameters) from a set of experimental measurements. Bayesian inference is theoretically more reasonable, computationally much simpler, and has advantages over traditional \(\chi^2\) fitting methods in quantifying uncertainty and revealing the underlying model parameters \cite{kennedy2001bayesian,van2021bayesian}. Hence, within the Bayesian framework, parameter uncertainty can be quantitatively estimated.\par

% 理论模型
This letter presents a Bayesian method to constrain quantitatively the extracted optimal incident energies (OIE) of heavy-ion fusion-evaporation reactions and the key parameters of the theoretical model from experimental ERCS data. Dubna provides more substantial experimental data for the reactions ${}^{48}\mathrm{Ca}+{}^{243}\mathrm{Am}$, ${}^{48}\mathrm{Ca}+{}^{242}\mathrm{Pu}$ and ${}^{48}\mathrm{Ca}+{}^{238}\mathrm{U}$, including the latest data based on the DGFRS-2 \cite{oganessian2022new,oganessian2022investigation}. This reaction is unique as the OIE can be approximately determined from experimental data, and its alpha decay lifetime and decay energy are relatively accurate. Therefore, their validity can be tested by studying this response through various theoretical models. \par

The model based on the dinuclear system (DNS) concept has been successfully used in investigating the mechanism of the synthesis of SHEs \cite{bao2015theoretical,huang2010competing,zhu2014production,zhang2018production,wen2020multinucleon,bao2015influence,guo2019effect}. In this study, we use the dinuclear system model (DNS-sysu) to investigate the ERCS.
The penetration probability is given by the well-known Hill-Wheeler formula \cite{hill1953nuclear}:
\begin{equation}
\begin{aligned}&T
(E_{\mathrm{c.m.}},J)\\&=\int\frac{f(B)dB}{1+\exp\left\{-\frac{2\pi}{\hbar\omega(J)}\left[E_{\mathrm{c.m.}}-B-\frac{\hbar^2}{2\mu R_{B}^2(J)}J(J+1)\right]\right\}}.
\end{aligned}
\end{equation}
Here, $\hbar\omega{(J)}$ is the width of the parabolic Coulomb barrier. \textit{B} is the height of the Coulomb barrier. $R_{B}(J)$ defines the position of the barrier, and $f(\textit{B})$ is the barrier distribution function which is taken as the asymmetric Gaussian form:\par
\begin{equation}
\begin{aligned}
f(B)=\frac1N\text{exp}\bigg[-\bigg(\frac{B-B_m}{\Delta_{1,2}}\bigg)^2\bigg],
\end{aligned}
\end{equation}
where $B_m=\frac{B_s+B_0}2$, $B_{0}$ is the height of the Coulomb barrier at waist-to-waist orientation and ground state deformation. $B_{s}$ is the minimum height of the Coulomb barrier with variance of dynamical deformation. The widths of the asymmetric Gaussian form is $\Delta_{1}=\frac{\Delta_{2}}2$ (for $B < B_m$ ) and $\Delta_{2}=\frac{B_0-B_s}4$ (for $B > B_m$ ). $N=\frac{\sqrt{\pi}(\Delta_1+\Delta_2)}{2}$ is the normalization constant.\par

After the projectile nucleus is captured by the target nucleus, fusion takes place forming the compound nucleus when the dinuclear system overcomes the inner fusion barrier \(B_\mathrm{fus}\). The more asymmetric configurations than those on the "B.G. line" are considered as the occurrence of fusion \cite{zhu2021unified}. Therefore, the probability of fusion can be expressed as:
\begin{equation}
\begin{aligned}
P_\mathrm{CN}(E_{\mathrm{c.m.}},J) = \sum_{\beta_2}\sum_{Z_1 = 1}^{Z_\mathrm{BG}}\sum_{N_1 = 1}^{N_\mathrm{BG}}P(Z_1,N_1,\beta_2,J,E_{\mathrm{c.m.}},\tau_{\mathrm{int}}),
\end{aligned}
\end{equation}
where the \(N_\mathrm{BG}\) and \(Z_\mathrm{BG}\) are the neutron number and charge number at the B.G. point depending on the dynamical deformation $\beta_2$, respectively. The time evolution of the distribution probability function $P(Z_1,N_1,\beta_2,J,E_{\mathrm{c.m.}},\tau_{\mathrm{int}})$ for fragment 1, with $Z_1$, $N_1$, $\beta_2$, and having incident energy $E_{\mathrm{c.m.}}$ at the interaction time \(\tau_{\mathrm{int}}\), is described by solving the master equations on the corresponding potential energy surface. The detailed description can be seen in Ref. \cite{zhu2021unified}.
 \(\tau_{\mathrm{int}}\) is the interaction time determined by using deflection function \cite{li1983distribution}. The quasi-fission has been included in our model self-consistently. In the DNS concept, the radial degree of freedom is frozen. Therefore, the reaction channels are defined on the mass and charge asymmetry degrees of freedom. Due to high inner fusion barrier, most of events end with quasi-fission, and the corresponding configurations are more symmetric than those on the "B.G. line" \cite{zhu2021unified}. The quasi-fission probability can be calculated as 1-\(P_\mathrm{CN}\).   

Both the calculation of the capture cross section and the fusion probability are mainly determined by the details of the nucleus-nucleus interaction potential $V$, which comprises the Coulomb potential \(V_\mathrm{C}\) and the nuclear potential \(V_\mathrm{N}\). \(V_\mathrm{C}\) takes the form in Ref. \cite{wong1973interaction}, and \(V_\mathrm{N}\) is written by the double-folding method \cite{adamian1996effective,zhu2023law}:
\begin{equation}
\begin{aligned}\label{eq4}
V_\mathrm{N}(\mathbf{R}) = & C_{0}\bigg\{\frac{F_{\mathrm{in}}-F_{\mathrm{ex}}}{\rho_{0}}\bigg[\int\rho_{1}^{2}(\mathbf{r})\rho_{2}(\mathbf{r}-\mathbf{R})d\mathbf{r}  \\
&+\left.\int\rho_1(\mathbf{r})\rho_2^2(\mathbf{r}-\mathbf{R})d\mathbf{r}\right] \\
&\left.+F_{\mathrm{ex}}\int\rho_{1}(\mathbf{r})\rho_{2}(\mathbf{r}-\mathbf{R})d\mathbf{r}\right\}.
\end{aligned}
\end{equation}
The nuclear density distribution functions \(\rho_1\) and \(\rho_2\) are given using the Woods-Saxon types \cite{wong1973interaction}:
\begin{equation}
\begin{aligned}
\rho_1(\mathbf{r},\theta_1) = \frac{\rho_0}{1+\exp[(\mathbf{r}-\Re_1(\theta_1))/a_1]},
\end{aligned}
\end{equation}
and
\begin{equation}
\begin{aligned}
\rho_2(\mathbf{r}-\mathbf{R},\theta_2) = \frac{\rho_0}{1+\exp[(\mid\mathbf{r}-\mathbf{R}\mid-\Re_2(\theta_2))/a_2]}.
\end{aligned}
\end{equation}
In the formula, \(\rho_0\) = 0.165 \({{\mathrm{fm}}^{-3}}\), \(\Re_i(\theta_i)\) is the surface radius of the $i$th nucleus. The diffuseness parameters for the light projectile and heavy target are denoted by $a_1$ and $a_2$, respectively. In this study, we assume that $a_1$ equals to $a_2$.\par 

The statistical approach is applied to calculate deexcitation probability \cite{xia2011systematic,zubov2002survival}. The Monte Carlo method is used to obtain the probabilities of all main possible decay  channels. 
% We based on the Monte Carlo method for calculating the decay probabilties in each channel\cite{zhu2021unified}.
In the \textit{i}th deexcitation step the probability of evaporating the neutron (n) channel can be written as
\begin{equation}
P_\text{n}(E_i^*)=\frac{\Gamma_\text{n}(E_i^*)}{\Gamma_{\text{tot}}(E_i^*)},
\end{equation}
where, \(\Gamma_\text{tot}=\Gamma_\text{n}+\Gamma_\text{p}+\Gamma_{\alpha}+\Gamma_{\gamma}+\Gamma_{\text{f}}\), which is addressed in detail in Ref. \cite{zhu2021unified,zhu2020selection}. The partial decay widths for the evaporation of neutron can be estimated by the Weisskopf-Ewing theory \cite{PhysRev.57.472}.
\begin{equation}\begin{aligned}
\Gamma_{\mathbf{n}}(E^{*},J)& =\frac{(2s_{\mathbf{n}}+1)m_{\mathbf{n}}}{\pi^2\hbar^2\rho(E^*,J)}  \\
&\times\int_{I_{\mathrm{n}}}\varepsilon\rho(E^{*}-B_{\mathrm{n}}-\varepsilon,J)\sigma_{\mathrm{inv}}(\varepsilon)d\varepsilon.
\end{aligned}\end{equation}
The level density is calculated as \cite{zagrebaev2015cross}:
\begin{equation}
\begin{gathered}
\rho(E^*,J)=K_{\mathrm{coll}}\frac{(2J+1)\sqrt{a_{\mathbf{n}}}}{24(E^*-\delta-{E_{\mathrm{rot}}})^2}\biggl(\frac{\hbar^2}\zeta\biggr)^{3/2}\\\times\exp[2\sqrt{a_{\mathbf{n}}(E^*-\delta-{E_{\mathrm{rot}}})}],
\end{gathered}
\end{equation} 
where $a_\mathrm{n} = \mathrm{A/12}$ $\mathrm{MeV}^{-1}$ is the level-density parameter of the neutron channel \cite{zhu2020selection}.

The fission decay width is usually calculated within the Bohr-Wheeler transition-state method \cite{bohr1939mechanism}.
\begin{equation}\begin{aligned}
\Gamma_{\mathbf{f}}(E^{*},J)& =\frac1{2\pi\rho_\mathrm{f}(E^*,J)}  \\
&\times\int_{I_\mathrm{f}}\frac{\rho_\mathrm{f}(E^*-B_\mathrm{f}-\varepsilon,J)d\varepsilon}{1+\exp[-2\pi(E^*-B_\mathrm{f}-\varepsilon)/\hbar\omega]}.
\end{aligned}\end{equation}
The level density in the fission decay width is calculated as \cite{zagrebaev2015cross}:
\begin{equation}
\begin{gathered}
\rho_{\mathbf{f}}(E^*,J)=K_{\mathrm{coll}}\frac{(2J+1)\sqrt{a_{\mathbf{f}}}}{24(E^*-\delta-{E_{\mathrm{rot}}})^2}\biggl(\frac{\hbar^2}\zeta\biggr)^{3/2}\\\times\exp[2\sqrt{a_{\mathbf{f}}(E^*-\delta-{E_{\mathrm{rot}}})}],
\end{gathered}
\end{equation} 
where \(a_{\mathrm{f}}\) is the level-density parameter of the fission channel. The fission barrier height \(B_\mathrm f(E^*)\) in \(\Gamma_\mathrm{f}\) is given as \cite{denisov2018calculation}:
\begin{equation}
\begin{aligned}
    B_\mathrm f(E^*) =-E_\mathrm{sh}^0\mathrm{e}^{-E^*/E_\mathrm{d}}.
\end{aligned}
\end{equation}

Here $E_\mathrm{sh}^0$ is the shell correction energy which is taken from Ref. \cite{moller1993nuclear}. $E_\mathrm{d}$ is the damping factor of the shell effects.\par
  
\begin{figure*}[http]
    \centering
    \includegraphics[width = 0.9\textwidth]{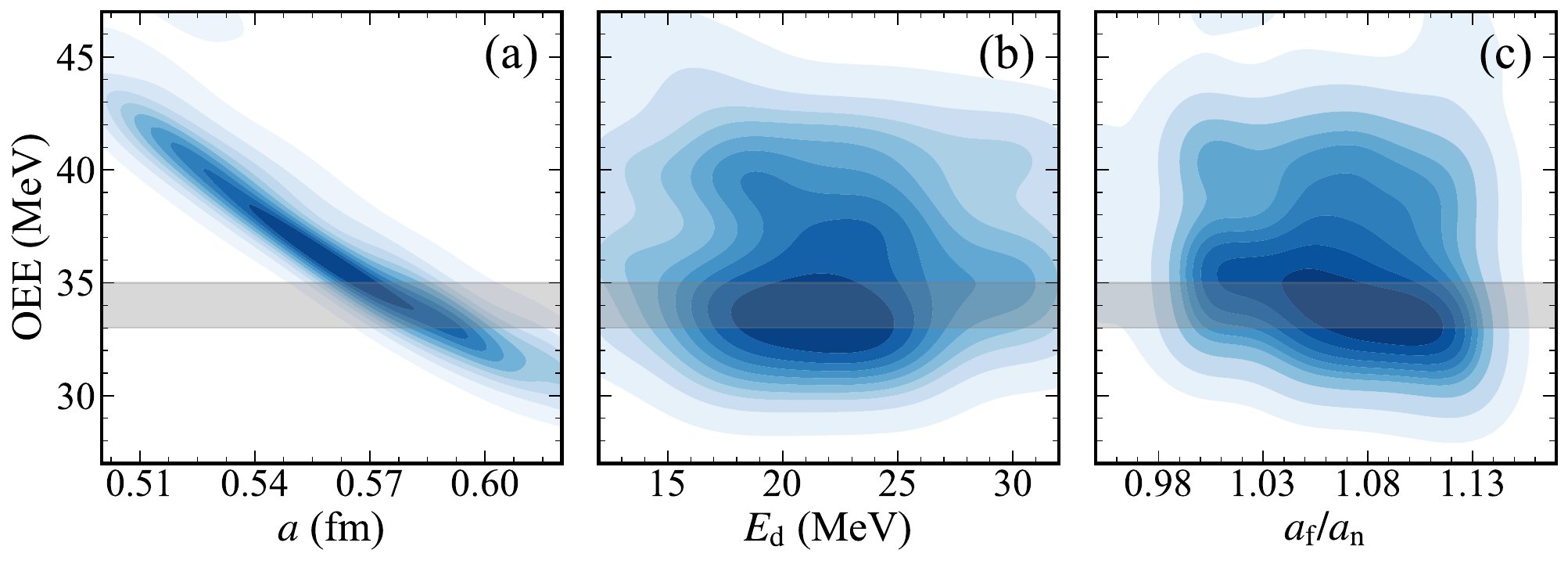}
    \caption{Probability density plots of (a) the diffusion parameter \textit{a}, (b) the damping factor $E_\mathrm{d}$, and (c) the level-density parameter ratio $a_\mathrm{f}/a_\mathrm{n}$ against OEE, respectively. The grey shaded region is OEE in the range $34 \pm1$ MeV. }
    % _{-0.5}^{+0.5}
    \label{fig1}
\end{figure*}

As described above, (i) in the capture process, the Coulomb barrier plays an important role and the uncertainty mainly results from the nuclear potential since the Coulomb potential can be accurately calculated. In nuclear potential as shown in Eq. (\ref{eq4}), the diffuseness parameter $a$ strongly influences the nuclear density distribution related to the range of the nuclear force. (ii) The inner fusion barrier determines the fusion probability in the DNS-sysu model, which also strongly depends on the nuclear interaction potential. (iii) In deexcitation process, the damping factor $E_\mathrm{d}$ and the level-density parameter ratio $a_\mathrm{f}/a_\mathrm{n}$ are the key parameters which determines the competition between neutron evaporation and the fission. For these three parameters in the process of synthesizing the SHEs, in this work, we get different combinations of \textit{a}, $E_\mathrm{d}$, and $a_\mathrm{f}/a_\mathrm{n}$ obtained by varying the values of each parameter. The results obtained by calculating these combinations within the DNS-sysu model are compared with experimental data to constrain the parameters. \par

The Bayes' theorem states \cite{sivia2006data}:
\begin{equation} 
    P(X|M) = \frac{P(M|X)P(X)}{\int P(M|X)P(X)dX}\propto P(M|X)P(X),
\end{equation}   
where $\textit{P}(X|M)$ is the posterior probability for the model parameter \(\textit{X}\) given the observed data set \(\textit{M}\). \(\textit{P}(M|X)\) is the likelihood function for a given theoretical model parameter \(\textit{X}\) to correctly predict the data \(\textit{M}\). \(\textit{P}(X)\) is the prior probability of parameter \(\textit{X}\) based on prior knowledge before any new observations are considered.\par

To avoid the complicated integral calculation, the posterior distribution is constructed by Markov chain Monte Carlo (MCMC) sampling \cite{foreman2013emcee}.
Based on the MCMC approach, we use the Metropolis-Hastings (MH) algorithm \cite{hastings1970monte} to generate a probabilistic sequence of Markov chains. This involves performing a weighted random walk in the parameter space. Beginning with an initial parameter distribution \(\textit{M}\) (t = 0), each step in the Markov chain results in a new set of parameters \(\textit{M}\) (t+1) through a random walk process. The decision to accept or reject the parameter set of the next moment is determined by the probability $P=\mathrm{min}\left(1,\frac{P(X^{t+1})P(M|X^{t+1})}{X^{t}P(M|X^{t})}\right)$. For the large sample requirements of MCMC sampling and alternative the computationally expensive DNS-sysu model, Gaussian process (GP) emulator \cite{ming2023deep} with Principal Component Analysis (PCA) \cite{mackiewicz1993principal} is utilised to interpolate the input-output behaviour of the model. The GP emulator serves a surrogate for the DNS-sysu model and provides a fast alternative to the computationally expensive the
DNS-sysu model.\par

There has been much discussion about the range of values for \textit{a}, $E_\mathrm{d}$, and $a_\mathrm{f}/a_\mathrm{n}$ \cite{ignatyuk1975fission,ignatyuk1979role,zhao2005entrance,adamian2000analysis,moller2016nuclear,zargini2023significance,zubov2003survivability}. The empirical range of the key parameters is taken as 0.50<\textit{a}<0.62 fm; 12<$E_\mathrm{d}$<32 MeV; and 0.95<$a_\mathrm{f}/a_\mathrm{n}$<1.2. We take 11, 10, and 7 points in their ranges, respectively, giving a total of 770 parameter sets. Based on the above parameter sets, we theoretically calculate the ERCS of the reactions ${}^{48}\mathrm{Ca}+{}^{243}\mathrm{Am}$, ${}^{48}\mathrm{Ca}+{}^{242}\mathrm{Pu}$, and ${}^{48}\mathrm{Ca}+{}^{238}\mathrm{U}$ for each set of parameters using the DNS-sysu model. Due to the fact that the OIE are crucial for conducting the experiments to hunt the new elements, we consider the experimental value of OIE as one important observable to constrain the range of the parameters. For the reactions ${}^{48}\mathrm{Ca}+{}^{243}\mathrm{Am}$, ${}^{48}\mathrm{Ca}+{}^{242}\mathrm{Pu}$, and ${}^{48}\mathrm{Ca}+{}^{238}\mathrm{U}$, the optimal excitation energy (OEE = OIE + $Q$) that we obtained from the experiment are about 34, 39, and 35 MeV, respectively, and the values of \(\textit{Q}\) calculated by using the Myers mass table are -165.99, -162.64, and -158.85 MeV, respectively \cite{myers1996nuclear}.\par

\begin{table}[]
\caption {\label{table}Mean, 1\({\sigma}\) confidence level, and 2\({\sigma}\) confidence level for \textit{a}, $E_\mathrm{d}$, and $a_\mathrm{f}/a_\mathrm{n}$ of the posterior distributions obtained from the Bayesian inference.}
\begin{tabular}{l@{\hspace{1em}}c@{\hspace{1em}}c@{\hspace{1em}}c}
\hline\hline
& mean   &  1${\sigma}$   &  2${\sigma}$  \\
\hline
$\text{\textit{a} (fm)} $      & 0.586   & 0.586 $\sim$ 0.587 & 0.585 $\sim$ 0.589\\
$\text{$E_\mathrm{d}$}~(\mathrm{MeV})$    & 25.65  & 23.93 $\sim$ 27.38 & 22.24 $\sim$ 29.08\\
$\text{$a_\mathrm{f}/a_\mathrm{n}$}$       & 1.081   & 1.070 $\sim$ 1.092 & 1.059 $\sim$ 1.102\\ 
\hline\hline                                                              
\end{tabular}
\end{table}

% 解释图1
Whether and how is the OIE affected by \textit{a}, $E_\mathrm{d}$, and $a_\mathrm{f}/a_\mathrm{n}$? We plot parameter-OEE probability density distributions in Fig. \ref{fig1}. The grey shaded region is plotted based on an experimentally measured OEE of 34 MeV for the reaction ${}^{48}\mathrm{Ca}+{}^{243}\mathrm{Am}$ with an error of 1 MeV. The above behavior also indicate that the OIE weakly depends on the  $E_\mathrm{d}$ and $a_\mathrm{f}/a_\mathrm{n}$. This is mainly because the diffusion parameter \textit{a} has a significant impact on the capture cross section, especially near the Coulomb barrier and consequently its variation has a considerable effect on ERCS in 2n and 3n channels, resulting in a relatively large uncertainty in the OEE. The results indicate that the OIE corresponding to the largest ERCS weakly depend on the fission process but closely associated with the capture and fusion processes. This finding would contribute significantly for theoretical estimation of the OIE. \par

Assuming that the prior of \textit{a}, $E_\mathrm{d}$, and $a_\mathrm{f}/a_\mathrm{n}$ are taken to be uniform distributed in their empirical range.
% the range of 0.5 - 0.62 fm, 12 - 32 MeV, and 0.95 - 1.2, respectively. 
The likelihood is given by:
\begin{equation}
    \begin{aligned}
        &P(M^{\exp}\mid X) \\
        & \propto\exp\left\{-\frac{1}{2}(M^{\text{emu}}-M^{\exp})^{T}
        \Sigma_\mathrm{M}^{-1}(M^{\text{emu}}-M^{\exp})\right\},
    \end{aligned}
\end{equation}
where the model parameter $X$ includes \textit{a}, $E_\mathrm{d}$, and $a_\mathrm{f}/a_\mathrm{n}$, \(M^{\mathrm{emu}}\) and $M^{\text{exp}}$ are the GP predictions and experimental values for the ERCS and OEE of the reaction ${}^{48}\mathrm{Ca}+{}^{243}\mathrm{Am}$, respectively. The covariance matrix $\Sigma_\mathrm{M}$ = \(\Sigma_{\mathrm{GP}}\) + diag(\(\sigma^2\)) includes the uncertainties and correlations of the emulator predictions via the GP covariance matrix \(\Sigma_{\mathrm{GP}}\), and the adopted error $\sigma$ which takes into account the deficiency of theoretical models and experimental error. The theoretical model error is considered to be 0, and the experimental error is determined by the ERCS error bars of the experiment.\par

% 解释表1、图2
Subsequently, the MCMC method of the MH algorithm to sample 1\(\times10^7\) sets in the parameter space. The first \(5\times10^6\) samples are treated as burn-in samples which are not used. By analysing the end-state information, table \ref{table} shows the results of the mean, 1\({\sigma}\) confidence level, and 2\({\sigma}\) confidence level for each parameter. We take the mean value of \textit{a}, $E_\mathrm{d}$, and $a_\mathrm{f}/a_\mathrm{n}$ as the optimal value, \textit{a} = 0.586 fm, $E_\mathrm{d}$ = 25.65 MeV, and $a_\mathrm{f}/a_\mathrm{n}$ = 1.081.\par

\begin{figure}[http]
    \centering
    \includegraphics[width = 0.48\textwidth]{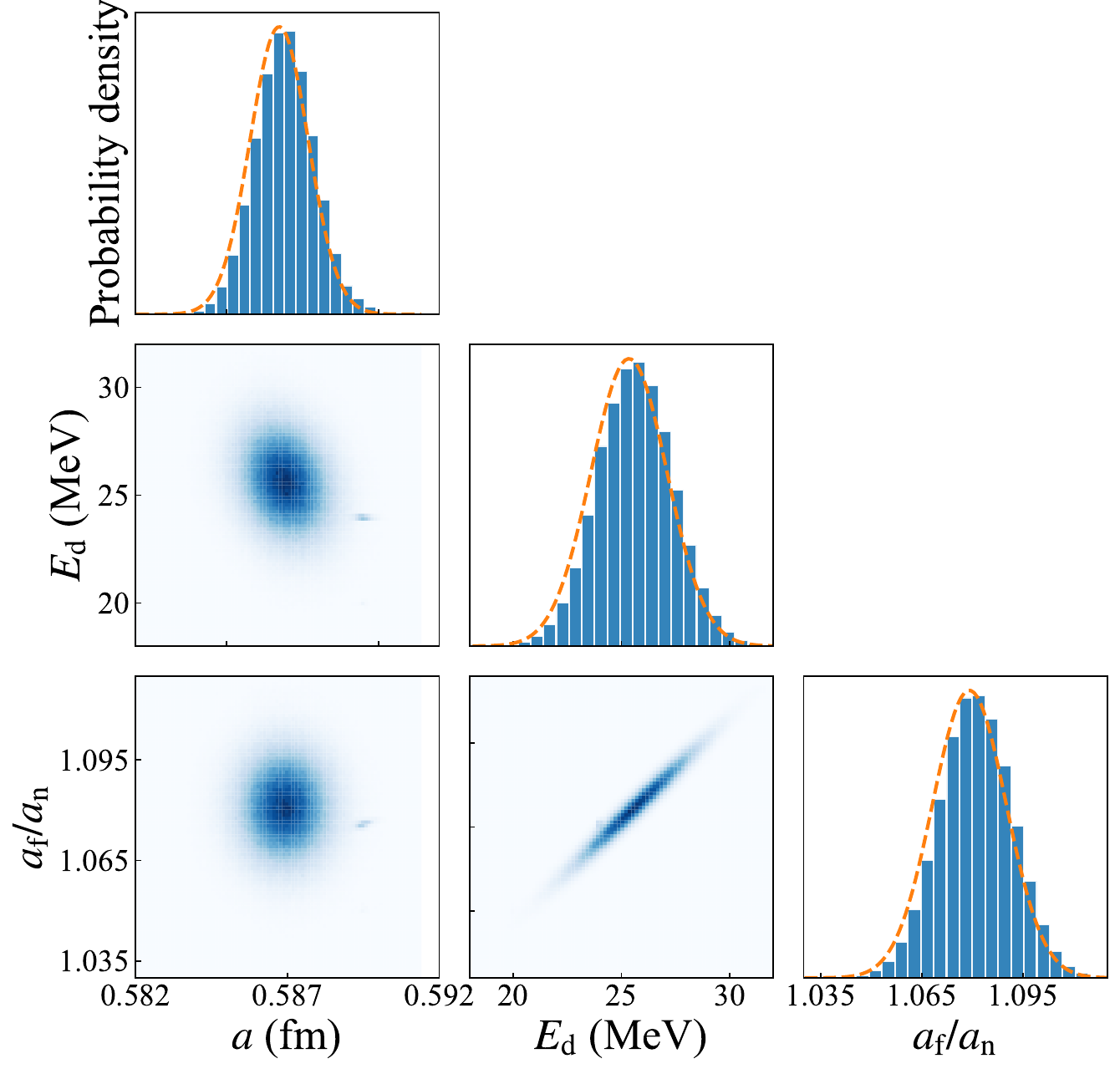}
    \caption{Univariate and bivariate marginal estimates of the posterior distribution for the diffusion parameter \textit{a}, the damping factor $E_\mathrm{d}$, and the level-density parameter ratio $a_\mathrm{f}/a_\mathrm{n}$.Orange dashed lines denote a Gaussian fit to the distribution.}
    \label{fig2}
\end{figure}

Bayesian inference results of \textit{a}, $E_\mathrm{d}$, and $a_\mathrm{f}/a_\mathrm{n}$ parameters are shown in Fig. \ref{fig2}. The univariate marginal posterior distributions of each key parameter shown as the histograms are plotted in the diagonal panels, while the posteriors of bivariate marginal  are shown in the off-diagonal panels. From these bivariate marginal posterior distributions, we see that
$E_\mathrm{d}$ with $a_\mathrm{f}/a_\mathrm{n}$ is constrained to narrow bars. The Pearson correlation coefficients of $E_\mathrm{d}$ with  $a_\mathrm{f}/a_\mathrm{n}$ is calculated to be 0.97, indicating a positive correlation between the parameter $E_\mathrm{d}$ and $a_\mathrm{f}/a_\mathrm{n}$. The results show that $E_\mathrm{d}$, and $a_\mathrm{f}/a_\mathrm{n}$ are strongly correlated. Hence, we would like to emphasize that it is not reasonable to simply follow the relationship between the key parameters independently to evaluate the uncertainty of the ERCS and OIE.
 
\begin{figure}[http]
    \centering
    \includegraphics[width = 0.4\textwidth]{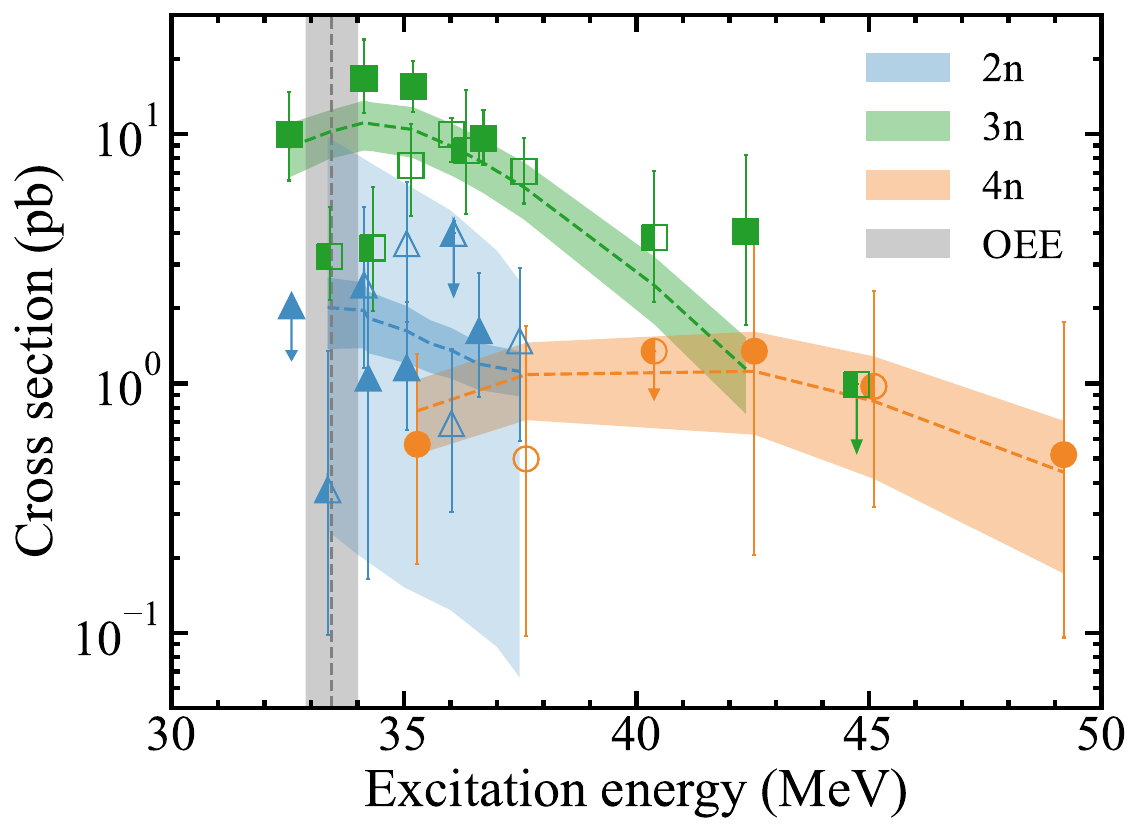}
    \caption{The dashed lines are the posterior median of the ERCS and OEE. Data shown by open, half-closed, and closed symbols are from \cite{khuyagbaatar2019fusion}, \cite{forsberg2016recoil}, and \cite{oganessian2022first}. The dark uncertainty bands correspond to the 2\({\sigma}\) confidence level constraint constructed from the posterior samples. The confidence level of 2n evaporation channel calculated without considering parameter correlations is shown as a light blue shade.}
    \label{fig3}
\end{figure}

\begin{figure*}[http]
    \centering
    \includegraphics[width = 1\textwidth]{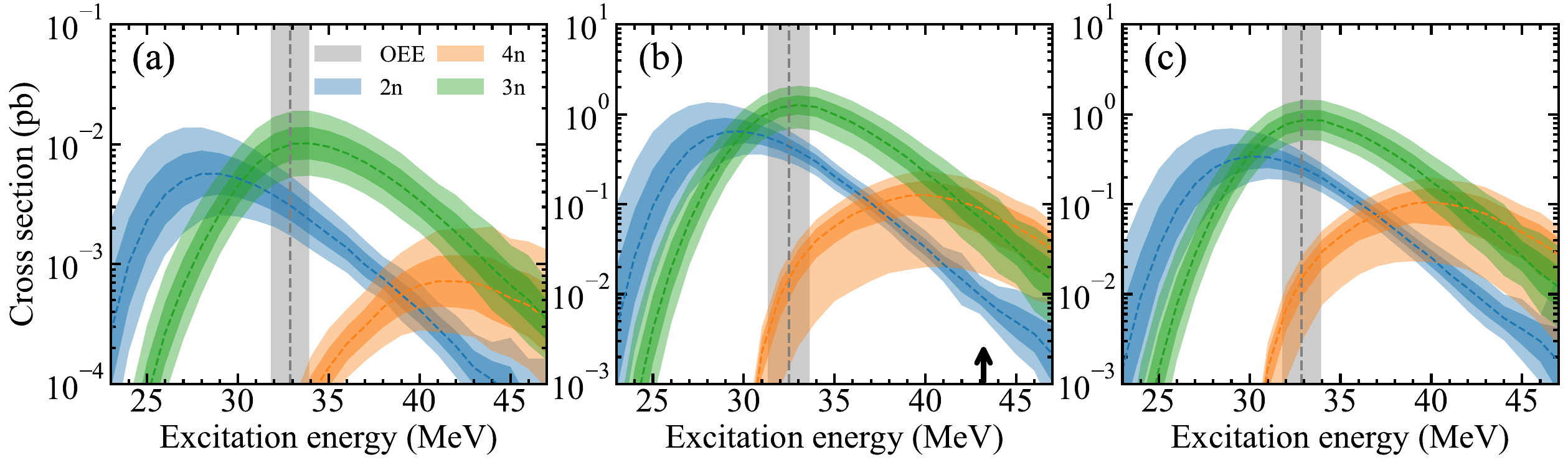}
    \caption{The ERCS and the OEE in reactions (a) \({}^{54}\mathrm{Cr}+{}^{243}\mathrm{Am}\), (b) \({}^{50}\mathrm{Ti}+{}^{249}\mathrm{Bk}\), and (c) \({}^{51}\mathrm{V}+{}^{248}\mathrm{Cm}\). The dark and light shaded bands represent the 1\({\sigma}\) and 2\({\sigma}\) confidence levels for the ERCS, respectively. The grey colored bands are the 2\({\sigma}\) confidence levels of the OEE. The dashed lines are obtained by taking the mean of ERCS and OEE. The black arrow in (b) denotes the incident energy used in the GSl for the reaction \({}^{50}\mathrm{Ti}+{}^{249}\mathrm{Bk}\) \cite{khuyagbaatar2020search}.}
    \label{fig4}
\end{figure*}

% 解释图3
By using Bayesian inference to propagate theoretical errors, we plot the 2\({\sigma}\) confidence levels of the ERCS and OEE for the reaction ${}^{48}\mathrm{Ca}+{}^{243}\mathrm{Am}$ as shown in Fig. \ref{fig3}. The confidence level of ERCS in each neutron evaporation channel is within one order of magnitude. The 2\({\sigma}\) confidence level range of the OIE is 199.5 $\sim$ 200.9 MeV, the mean value is 200.2 MeV, which is close to the value of 198 MeV presented in Ref. \cite{zhu2023law}. These suggest a reasonable agreement between the ERCS and OEE calculated within the Bayesian statistical framework using the DNS-sysu model. Deviations are within acceptable error margins. The results offer a quantitative statement of the uncertainty introduced by the key parameters into the theoretical model. Note that the light blue shade in the confidence level of the 2n channel is relatively large, approximately two orders of magnitude, indicating that the uncertainty of ERCS obtained without considering the correlation between the parameters is strongly overestimated. \par

% 过渡到119
In order to further investigate ERCS and OIE of synthesizing SHEs with Z = 119, as well as to quantitatively give an estimate of the uncertainty on the DNS-sysu model prediction for Z = 119. 
Initially, we calculate the $Q$ values for the reactions \({}^{54}\mathrm{Cr}+{}^{243}\mathrm{Am}\), \({}^{50}\mathrm{Ti}+{}^{249}\mathrm{Bk}\), and \({}^{51}\mathrm{V}+{}^{248}\mathrm{Cm}\) as -206.28, -191.47, and -195.34 MeV, respectively.
Subsequently, we randomly selected 50 parameter sets based on the posterior distributions of the key parameters in Fig. \ref{fig2}. We obtained the confidence level of the ERCS and OEE for the different reaction systems. Fig. \ref{fig4}(a), (b), and (c) show the 1\({\sigma}\) and 2\({\sigma}\) confidence levels of the ERCS and OEE for the reactions ${}^{54}\mathrm{Cr}+{}^{243}\mathrm{Am}$, ${}^{50}\mathrm{Ti}+{}^{249}\mathrm{Bk}$, and ${}^{51}\mathrm{V}+{}^{248}\mathrm{Cm}$, respectively. To verify the reliability, we repeat the process for another 50 sets of parameters, and the close results are noticed as before, indicating that these 50 parameter sets are a good representation of the posterior distribution of the parameters. In Fig. \ref{fig4}(a), (b), and (c), the gray shaded regions represent the confidence levels of the OEE, corresponding to 238.1 $\sim$ 240.2 MeV, 222.8 $\sim$ 225.1 MeV, and 227.1 $\sim$ 229.3 MeV of OIE for the reactions ${}^{54}\mathrm{Cr}+{}^{243}\mathrm{Am}$, ${}^{50}\mathrm{Ti}+{}^{249}\mathrm{Bk}$, and ${}^{51}\mathrm{V}+{}^{248}\mathrm{Cm}$, respectively. 
Khuyagbaatar et al. investigated the reaction ${}^{50}\mathrm{Ti}+{}^{249}\mathrm{Bk}$ to produce the SHE Z = 119 at $E_\mathrm{c.m.}$ around 234.4 MeV \cite{khuyagbaatar2020search}. However, no event was observed. In Fig. 4(b), the black arrow denotes excitation energy corresponding to the incident energy used in the GSI. At this incident energy ERCS = $158_{-140}^{+102}$ fb from our calculation under the 2\({\sigma}\) confidence level. The calculated lower limit of the ERCS is 18 fb, which is much lower than cross-section sensitivity of 65 fb reached in GSI-TASCA experiments \cite{khuyagbaatar2020search}. Actually, the correct incident energy is essential. From our calculation, the incident energy used in GSI is much higher than optimal one, which could be the main reason for failure of synthesizing 119 element. In our theoretical calculations, OIE = 223.95 MeV, and the corresponding ERCS is in the range of 0.958 to 2.76 pb under the 2\({\sigma}\) confidence level. Meanwhile, from Fig. \ref{fig4}, it can be seen that the confidence level of ERCS fluctuates within an order of magnitude, without straightforwardly accumulating the uncertainties of individual parameters to such a significant extent. Furthermore, given that the combined effect of these parameters on the ERCS and OIE are not expected to be large, this is favourable for theoretical predictions of the synthesis of SHEs.\par

% \section{Summary and conclusions}
In summary, we have presented a comprehensive application of Bayesian inference method to the calculation and propagation of the key parameters' uncertainties in the DNS-sysu model. An interesting observation reveals that the OIE associated with the highest ERCS, exhibits weak dependence on the fission process. Through the utilization of the latest dataset concerning the reactions ${}^{48}\mathrm{Ca}+{}^{243}\mathrm{Am}$, ${}^{48}\mathrm{Ca}+{}^{242}\mathrm{Pu}$ and ${}^{48}\mathrm{Ca}+{}^{238}\mathrm{U}$ from the Dubna Laboratory, we quantitatively give the confidence levels for the parameters in the DNS-sysu model, $\textit{a} = 0.586_{-0.002}^{+0.002}$ fm, $E_\mathrm{d} = 25.65_{-3.41}^{+3.43}$ MeV, and $a_{\mathrm{f}}/a_{\mathrm{n}} = 1.081_{-0.021}^{+0.021}$ at 2\({\sigma}\) confidence level. Importantly, the strong correlations among $a$, $E_\mathrm{d}$, and $a_\mathrm{f}/a_\mathrm{n}$ are noticed, which indicates that the uncertainty propagation of the the parameters in fusion-evaporation reactions is not mutually independent. Neglecting the uncertainty estimation of their correlations results in a significant overestimation. We have accounted for their correlations. The results illustrate that the confidence levels for the ERCS and OIE are within reasonable experimental errors. Finally, we predict the 1\({\sigma}\) and 2\({\sigma}\) confidence levels for the ERCS and OIE of the reactions ${}^{54}\mathrm{Cr}+{}^{243}\mathrm{Am}$, ${}^{50}\mathrm{Ti}+{}^{249}\mathrm{Bk}$, and ${}^{51}\mathrm{V}+{}^{248}\mathrm{Cm}$.\par

We conduct a rigorous analysis of the uncertainty of key parameters in the DNS-sysu model. By applying the Bayesian framework to propagate theoretical statistical uncertainties in predictions of the OIE, the current results are statistically more reliable. Beyond the uncertainties from the key parameters, there are still several factors influence the reliability of the ERCS predictions and need long-term development. (i) The nuclides masses and the fission barriers are important inputs, which depends on the theoretical extrapolation. Therefore, weakening the model dependence and the reliable calculations of these quantities are very important for predicting the ERCS. (ii) As we known that the ERCS predictions from different models (usually different physical pictures of fusion) show large discrepancies. Therefore, it is necessary to find out and combine the major degrees of freedom for describing the fusion. In the DNS concept, the radial evolution is frozen. In the future, we would employ our recently developed model \cite{zhu2024new} in study the synthesis of SHN, in which the mass asymmetry and radial degrees of freedom are both considered. (iii) The more reasonable description of shell effects on level density and their degradation with excitation energy in the survival process should be further developed, especially for the Z=119 and 120 SHEs around possible magic numbers.

\section*{Acknowledgements}
The authors would like to thank Feng-Shou Zhang, Shan-Gui Zhou, Zai-Guo Gan, Cheng-Jian Lin, Hong-Fei Zhang, Ning Wang, Nan Wang, Jun-Chen Pei, Hui-Min Jia, Hua-Bin Yang, Bing Wang, Xiao-Jun Bao, and Yong-Jia Wang for helpful discussion and suggestions. This work was supported by the National Natural Science Foundation of China under Grants No. 12075327 and 12335008; The Open Project of Guangxi Key Laboratory of Nuclear Physics and Nuclear Technology under Grant No. NLK2022-01; Fundamental Research Funds for the Central Universities, Sun Yat-sen University under Grant No. 23lgbj003. The work is supported in part by National Key R$\text{\&}$D Program of China
(2023YFA1606402).

\bibliographystyle{ieeetr} 
\bibliography{main}

\begin{thebibliography}{10}

\bibitem{ohrstrom2016names}
L.~{\"O}hrstr{\"o}m and J.~Reedijk, ``Names and symbols of the elements with
  atomic numbers 113, 115, 117 and 118 (iupac recommendations 2016),'' {\em
  Pure and Applied Chemistry}, vol.~88, no.~12, pp.~1225--1229, 2016.

\bibitem{oganessian2015super}
Y.~T. Oganessian and V.~Utyonkov, ``Super-heavy element research,'' {\em
  Reports on Progress in Physics}, vol.~78, no.~3, p.~036301, 2015.

\bibitem{RevModPhys.91.011001}
S.~A. Giuliani, Z.~Matheson, W.~Nazarewicz, E.~Olsen, P.-G. Reinhard,
  J.~Sadhukhan, B.~Schuetrumpf, N.~Schunck, and P.~Schwerdtfeger, ``Colloquium:
  Superheavy elements: Oganesson and beyond,'' {\em Rev. Mod. Phys.}, vol.~91,
  p.~011001, Jan 2019.

\bibitem{morita2012new}
K.~Morita, K.~Morimoto, D.~Kaji, H.~Haba, K.~Ozeki, Y.~Kudou, T.~Sumita,
  Y.~Wakabayashi, A.~Yoneda, K.~Tanaka, {\em et~al.}, ``New result in the
  production and decay of an isotope, 278113, of the 113th element,'' {\em
  Journal of the Physical Society of Japan}, vol.~81, no.~10, p.~103201, 2012.

\bibitem{RevModPhys.72.733}
S.~Hofmann and G.~M\"unzenberg, ``The discovery of the heaviest elements,''
  {\em Rev. Mod. Phys.}, vol.~72, pp.~733--767, Jul 2000.

\bibitem{hofmann2016remarks}
S.~Hofmann, S.~Heinz, R.~Mann, J.~Maurer, G.~M{\"u}nzenberg, S.~Antalic,
  W.~Barth, L.~Dahl, K.~Eberhardt, R.~Grzywacz, {\em et~al.}, ``Remarks on the
  fission barriers of super-heavy nuclei,'' {\em The European Physical Journal
  A}, vol.~52, pp.~1--12, 2016.

\bibitem{khuyagbaatar2020search}
J.~Khuyagbaatar, A.~Yakushev, C.~E. D{\"u}llmann, D.~Ackermann, L.-L.
  Andersson, M.~Asai, M.~Block, R.~Boll, H.~Brand, D.~Cox, {\em et~al.},
  ``Search for elements 119 and 120,'' {\em Physical Review C}, vol.~102,
  no.~6, p.~064602, 2020.

\bibitem{oganessian2009attempt}
Y.~T. Oganessian, V.~Utyonkov, Y.~V. Lobanov, F.~S. Abdullin, A.~Polyakov,
  R.~Sagaidak, I.~Shirokovsky, Y.~S. Tsyganov, A.~Voinov, A.~Mezentsev, {\em
  et~al.}, ``Attempt to produce element 120 in the pu 244+ fe 58 reaction,''
  {\em Physical review C}, vol.~79, no.~2, p.~024603, 2009.

\bibitem{tanaka2022probing}
M.~Tanaka, P.~Brionnet, M.~Du, J.~Ezold, K.~Felker, B.~J. Gall, S.~Go, R.~K.
  Grzywacz, H.~Haba, K.~Hagino, {\em et~al.}, ``Probing optimal reaction energy
  for synthesis of element 119 from 51v+ 248cm reaction with quasielastic
  barrier distribution measurement,'' {\em Journal of the Physical Society of
  Japan}, vol.~91, no.~8, p.~084201, 2022.

\bibitem{li2023possibility}
J.-X. Li, H.-F. Zhang, {\em et~al.}, ``Possibility to synthesize z> 118
  superheavy nuclei with cr 54 projectiles,'' {\em Physical Review C},
  vol.~108, no.~4, p.~044604, 2023.

\bibitem{hagino2018hot}
K.~Hagino, ``Hot fusion reactions with deformed nuclei for synthesis of
  superheavy nuclei: An extension of the fusion-by-diffusion model,'' {\em
  Physical Review C}, vol.~98, no.~1, p.~014607, 2018.

\bibitem{adamian1998fusion}
G.~Adamian, N.~Antonenko, W.~Scheid, and V.~Volkov, ``Fusion cross sections for
  superheavy nuclei in the dinuclear system concept,'' {\em Nuclear Physics A},
  vol.~633, no.~3, pp.~409--420, 1998.

\bibitem{lu2016synthesis}
H.~L{\"u}, D.~Boilley, Y.~Abe, and C.~Shen, ``Synthesis of superheavy elements:
  Uncertainty analysis to improve the predictive power of reaction models,''
  {\em Physical Review C}, vol.~94, no.~3, p.~034616, 2016.

\bibitem{loveland2015experimentalist}
W.~Loveland, ``An experimentalist’s view of the uncertainties in
  understanding heavy element synthesis,'' {\em The European Physical Journal
  A}, vol.~51, pp.~1--7, 2015.

\bibitem{wang2012theoretical}
N.~Wang, E.-G. Zhao, W.~Scheid, S.-G. Zhou, {\em et~al.}, ``Theoretical study
  of the synthesis of superheavy nuclei with z= 119 and 120 in heavy-ion
  reactions with trans-uranium targets,'' {\em Physical Review C}, vol.~85,
  no.~4, p.~041601, 2012.

\bibitem{kayumov2022analysis}
B.~Kayumov, O.~Ganiev, A.~Nasirov, and G.~Yuldasheva, ``Analysis of the fusion
  mechanism in the synthesis of superheavy element 119 via the cr 54+ am 243
  reaction,'' {\em Physical Review C}, vol.~105, no.~1, p.~014618, 2022.

\bibitem{siwek2012predictions}
K.~Siwek-Wilczy{\'n}ska, T.~Cap, M.~Kowal, A.~Sobiczewski, and
  J.~Wilczy{\'n}ski, ``Predictions of the fusion-by-diffusion model for the
  synthesis cross sections of z= 114--120 elements based on
  macroscopic-microscopic fission barriers,'' {\em Physical Review C}, vol.~86,
  no.~1, p.~014611, 2012.

\bibitem{zhu2014production}
L.~Zhu, W.-J. Xie, and F.-S. Zhang, ``Production cross sections of superheavy
  elements z= 119 and 120 in hot fusion reactions,'' {\em Physical Review C},
  vol.~89, no.~2, p.~024615, 2014.

\bibitem{deng2023examination}
X.-Q. Deng, S.-G. Zhou, {\em et~al.}, ``Examination of promising reactions with
  am 241 and cm 244 targets for the synthesis of new superheavy elements within
  the dinuclear system model with a dynamical potential energy surface,'' {\em
  Physical Review C}, vol.~107, no.~1, p.~014616, 2023.

\bibitem{PhysRevC.109.014622}
M.-H. Zhang, Y.-H. Zhang, Y.~Zou, C.~Wang, L.~Zhu, and F.-S. Zhang,
  ``Predictions of synthesizing elements with $z=119$ and 120 in fusion
  reactions,'' {\em Phys. Rev. C}, vol.~109, p.~014622, Jan 2024.

\bibitem{PhysRevC.94.034616}
H.~L\"u, D.~Boilley, Y.~Abe, and C.~Shen, ``Synthesis of superheavy elements:
  Uncertainty analysis to improve the predictive power of reaction models,''
  {\em Phys. Rev. C}, vol.~94, p.~034616, Sep 2016.

\bibitem{he2023machine}
W.~He, Q.~Li, Y.~Ma, Z.~Niu, J.~Pei, and Y.~Zhang, ``Machine learning in
  nuclear physics at low and intermediate energies,'' {\em Science China
  Physics, Mechanics \& Astronomy}, vol.~66, no.~8, p.~282001, 2023.

\bibitem{he2023high}
W.-B. He, Y.-G. Ma, L.-G. Pang, H.-C. Song, and K.~Zhou, ``High-energy nuclear
  physics meets machine learning,'' {\em Nuclear Science and Techniques},
  vol.~34, no.~6, p.~88, 2023.

\bibitem{boehnlein2022colloquium}
A.~Boehnlein, M.~Diefenthaler, N.~Sato, M.~Schram, V.~Ziegler, C.~Fanelli,
  M.~Hjorth-Jensen, T.~Horn, M.~P. Kuchera, D.~Lee, {\em et~al.}, ``Colloquium:
  Machine learning in nuclear physics,'' {\em Reviews of Modern Physics},
  vol.~94, no.~3, p.~031003, 2022.

\bibitem{wang2023machine}
Y.~Wang and Q.~Li, ``Machine learning transforms the inference of the nuclear
  equation of state,'' {\em Frontiers of Physics}, vol.~18, no.~6, p.~64402,
  2023.

\bibitem{ma2023phase}
Y.-G. Ma, L.-G. Pang, R.~Wang, and K.~Zhou, ``Phase transition study meets
  machine learning,'' {\em Chinese Physics Letters}, vol.~40, no.~12,
  p.~122101, 2023.

\bibitem{huth2022constraining}
S.~Huth, P.~T. Pang, I.~Tews, T.~Dietrich, A.~Le~F{\`e}vre, A.~Schwenk,
  W.~Trautmann, K.~Agarwal, M.~Bulla, M.~W. Coughlin, {\em et~al.},
  ``Constraining neutron-star matter with microscopic and macroscopic
  collisions,'' {\em Nature}, vol.~606, no.~7913, pp.~276--280, 2022.

\bibitem{qiu2024bayesian}
M.~Qiu, B.-J. Cai, L.-W. Chen, C.-X. Yuan, and Z.~Zhang, ``Bayesian model
  averaging for nuclear symmetry energy from effective proton-neutron chemical
  potential difference of neutron-rich nuclei,'' {\em Physics Letters B},
  vol.~849, p.~138435, 2024.

\bibitem{li2024importance}
Z.~Li, Z.~Gao, L.~Liu, Y.~Wang, L.~Zhu, and Q.~Li, ``Importance of physical
  information on the prediction of heavy-ion fusion cross sections with machine
  learning,'' {\em Physical Review C}, vol.~109, no.~2, p.~024604, 2024.

\bibitem{xu2021bayesian}
J.~Xu, Z.~Zhang, and B.-A. Li, ``Bayesian uncertainty quantification for
  nuclear matter incompressibility,'' {\em Physical Review C}, vol.~104, no.~5,
  p.~054324, 2021.

\bibitem{wang2019bayesian}
Z.-A. Wang, J.~Pei, Y.~Liu, and Y.~Qiang, ``Bayesian evaluation of incomplete
  fission yields,'' {\em Physical review letters}, vol.~123, no.~12, p.~122501,
  2019.

\bibitem{kennedy2001bayesian}
M.~C. Kennedy and A.~O'Hagan, ``Bayesian calibration of computer models,'' {\em
  Journal of the Royal Statistical Society: Series B (Statistical
  Methodology)}, vol.~63, no.~3, pp.~425--464, 2001.

\bibitem{van2021bayesian}
R.~van~de Schoot, S.~Depaoli, R.~King, B.~Kramer, K.~M{\"a}rtens, M.~G.
  Tadesse, M.~Vannucci, A.~Gelman, D.~Veen, J.~Willemsen, {\em et~al.},
  ``Bayesian statistics and modelling,'' {\em Nature Reviews Methods Primers},
  vol.~1, no.~1, p.~1, 2021.

\bibitem{oganessian2022new}
Y.~T. Oganessian, V.~Utyonkov, N.~Kovrizhnykh, F.~S. Abdullin, S.~Dmitriev,
  A.~Dzhioev, D.~Ibadullayev, M.~Itkis, A.~Karpov, D.~Kuznetsov, {\em et~al.},
  ``New isotope mc 286 produced in the am 243+ ca 48 reaction,'' {\em Physical
  Review C}, vol.~106, no.~6, p.~064306, 2022.

\bibitem{oganessian2022investigation}
Y.~T. Oganessian, V.~Utyonkov, D.~Ibadullayev, F.~S. Abdullin, S.~Dmitriev,
  M.~Itkis, A.~Karpov, N.~Kovrizhnykh, D.~Kuznetsov, O.~Petrushkin, {\em
  et~al.}, ``Investigation of ca 48-induced reactions with pu 242 and u 238
  targets at the jinr superheavy element factory,'' {\em Physical Review C},
  vol.~106, no.~2, p.~024612, 2022.

\bibitem{bao2015theoretical}
X.~Bao, Y.~Gao, J.~Li, and H.~Zhang, ``Theoretical study of the synthesis of
  superheavy nuclei using radioactive beams,'' {\em Physical Review C},
  vol.~91, no.~6, p.~064612, 2015.

\bibitem{huang2010competing}
M.~Huang, Z.~Gan, X.~Zhou, J.~Li, and W.~Scheid, ``Competing fusion and
  quasifission reaction mechanisms in the production of superheavy nuclei,''
  {\em Physical Review C}, vol.~82, no.~4, p.~044614, 2010.

\bibitem{zhang2018production}
F.-S. Zhang, C.~Li, L.~Zhu, and P.~Wen, ``Production cross sections for exotic
  nuclei with multinucleon transfer reactions,'' {\em Frontiers of Physics},
  vol.~13, pp.~1--16, 2018.

\bibitem{wen2020multinucleon}
P.~Wen, A.~Nasirov, C.~Lin, and H.~Jia, ``Multinucleon transfer reaction from
  view point of dynamical dinuclear system method,'' {\em Journal of Physics G:
  Nuclear and Particle Physics}, vol.~47, no.~7, p.~075106, 2020.

\bibitem{bao2015influence}
X.~Bao, Y.~Gao, J.~Li, H.~Zhang, {\em et~al.}, ``Influence of the nuclear
  dynamical deformation on production cross sections of superheavy nuclei,''
  {\em Physical Review C}, vol.~91, no.~1, p.~011603, 2015.

\bibitem{guo2019effect}
S.~Q. Guo, X.~J. Bao, H.~F. Zhang, J.~Q. Li, and N.~Wang, ``Effect of dynamical
  deformation on the production distribution in multinucleon transfer
  reactions,'' {\em Physical Review C}, vol.~100, no.~5, p.~054616, 2019.

\bibitem{hill1953nuclear}
D.~L. Hill and J.~A. Wheeler, ``Nuclear constitution and the interpretation of
  fission phenomena,'' {\em Physical Review}, vol.~89, no.~5, p.~1102, 1953.

\bibitem{zhu2021unified}
L.~Zhu and J.~Su, ``Unified description of fusion and multinucleon transfer
  processes within the dinuclear system model,'' {\em Physical Review C},
  vol.~104, no.~4, p.~044606, 2021.

\bibitem{li1983distribution}
J.~Li and G.~Wolschin, ``Distribution of the dissipated angular momentum in
  heavy-ion collisions,'' {\em Physical Review C}, vol.~27, no.~2, p.~590,
  1983.

\bibitem{wong1973interaction}
C.~Wong, ``Interaction barrier in charged-particle nuclear reactions,'' {\em
  Physical Review Letters}, vol.~31, no.~12, p.~766, 1973.

\bibitem{adamian1996effective}
G.~Adamian, N.~Antonenko, R.~Jolos, S.~Ivanova, and O.~Melnikova, ``Effective
  nucleus-nucleus potential for calculation of potential energy of a dinuclear
  system,'' {\em International Journal of Modern Physics E}, vol.~5, no.~01,
  pp.~191--216, 1996.

\bibitem{zhu2023law}
L.~Zhu, ``Law of optimal incident energy for synthesizing superheavy elements
  in hot fusion reactions,'' {\em Physical Review Research}, vol.~5, no.~2,
  p.~L022030, 2023.

\bibitem{xia2011systematic}
C.~Xia, B.~Sun, E.~Zhao, and S.~Zhou, ``Systematic study of survival
  probability of excited superheavy nuclei,'' {\em Science China Physics,
  Mechanics and Astronomy}, vol.~54, pp.~109--113, 2011.

\bibitem{zubov2002survival}
A.~Zubov, G.~Adamian, N.~Antonenko, S.~Ivanova, and W.~Scheid, ``Survival
  probability of superheavy nuclei,'' {\em Physical Review C}, vol.~65, no.~2,
  p.~024308, 2002.

\bibitem{zhu2020selection}
L.~Zhu, ``Selection of projectiles for producing trans-uranium nuclei in
  transfer reactions within the improved dinuclear system model,'' {\em Journal
  of Physics G: Nuclear and Particle Physics}, vol.~47, no.~6, p.~065107, 2020.

\bibitem{PhysRev.57.472}
V.~F. Weisskopf and D.~H. Ewing, ``On the yield of nuclear reactions with heavy
  elements,'' {\em Phys. Rev.}, vol.~57, pp.~472--485, Mar 1940.

\bibitem{zagrebaev2015cross}
V.~Zagrebaev and W.~Greiner, ``Cross sections for the production of superheavy
  nuclei,'' {\em Nuclear Physics A}, vol.~944, pp.~257--307, 2015.

\bibitem{bohr1939mechanism}
N.~Bohr and J.~A. Wheeler, ``The mechanism of nuclear fission,'' {\em Physical
  Review}, vol.~56, no.~5, p.~426, 1939.

\bibitem{denisov2018calculation}
V.~Y. Denisov and I.~Y. Sedykh, ``Calculation of the fission width of an
  excited nucleus with the fission barrier dependent on excitation energy,''
  {\em Physical Review C}, vol.~98, no.~2, p.~024601, 2018.

\bibitem{moller1993nuclear}
P.~M{\"o}ller, J.~Nix, W.~Myers, and W.~Swiatecki, ``Nuclear ground-state
  masses and deformations,'' {\em arXiv preprint nucl-th/9308022}, 1993.

\bibitem{sivia2006data}
D.~Sivia and J.~Skilling, {\em Data analysis: a Bayesian tutorial}.
\newblock OUP Oxford, 2006.

\bibitem{foreman2013emcee}
D.~Foreman-Mackey, D.~W. Hogg, D.~Lang, and J.~Goodman, ``emcee: the mcmc
  hammer,'' {\em Publications of the Astronomical Society of the Pacific},
  vol.~125, no.~925, p.~306, 2013.

\bibitem{hastings1970monte}
W.~K. Hastings, ``Monte carlo sampling methods using markov chains and their
  applications,'' 1970.

\bibitem{ming2023deep}
D.~Ming, D.~Williamson, and S.~Guillas, ``Deep gaussian process emulation using
  stochastic imputation,'' {\em Technometrics}, vol.~65, no.~2, pp.~150--161,
  2023.

\bibitem{mackiewicz1993principal}
A.~Ma{\'c}kiewicz and W.~Ratajczak, ``Principal components analysis (pca),''
  {\em Computers \& Geosciences}, vol.~19, no.~3, pp.~303--342, 1993.

\bibitem{ignatyuk1975fission}
A.~Ignatyuk, M.~Itkis, V.~Okolovich, G.~Smirenkin, and A.~Tishin, ``Fission of
  pre-actinide nuclei. excitation functions for the ($\alpha$, f) reaction,''
  {\em Yadernaya Fizika}, vol.~21, no.~6, pp.~1185--1205, 1975.

\bibitem{ignatyuk1979role}
A.~Ignatyuk, K.~Istekov, and G.~Smirenkin, ``Role of collective effects in the
  systematics of nuclear level densities,'' {\em Sov. J. Nucl. Phys.(Engl.
  Transl.);(United States)}, vol.~29, no.~4, 1979.

\bibitem{zhao2005entrance}
F.~Zhao-Qing, J.~Gen-Ming, F.~Fen, Z.~Feng-Shou, J.~Fei, H.~Xi, H.~Rong-Jiang,
  L.~Wen-Fei, and L.~Jun-Qing, ``Entrance channel dependence of production
  cross sections of superheavy nuclei in cold fusion reactions,'' {\em Chinese
  Physics Letters}, vol.~22, no.~4, p.~846, 2005.

\bibitem{adamian2000analysis}
G.~Adamian, N.~Antonenko, S.~Ivanova, and W.~Scheid, ``Analysis of survival
  probability of superheavy nuclei,'' {\em Physical Review C}, vol.~62, no.~6,
  p.~064303, 2000.

\bibitem{moller2016nuclear}
P.~M{\"o}ller, A.~J. Sierk, T.~Ichikawa, and H.~Sagawa, ``Nuclear ground-state
  masses and deformations: Frdm (2012),'' {\em Atomic Data and Nuclear Data
  Tables}, vol.~109, pp.~1--204, 2016.

\bibitem{zargini2023significance}
R.~Zargini and S.~Seyyedi, ``Significance of the compound nucleus surface
  energy coefficients in the synthesis of the superheavy nuclei with z=
  112-120,'' {\em arXiv preprint arXiv:2312.17567}, 2023.

\bibitem{zubov2003survivability}
A.~Zubov, G.~Adamian, N.~Antonenko, S.~Ivanova, and W.~Scheid, ``Survivability
  of excited superheavy nuclei,'' {\em Physics of Atomic Nuclei}, vol.~66,
  pp.~218--232, 2003.

\bibitem{myers1996nuclear}
W.~Myers and W.~Swiatecki, ``Nuclear properties according to the thomas-fermi
  model,'' {\em Nuclear Physics A}, vol.~601, no.~2, pp.~141--167, 1996.

\bibitem{khuyagbaatar2019fusion}
J.~Khuyagbaatar, A.~Yakushev, C.~E. D{\"u}llmann, D.~Ackermann, L.-L.
  Andersson, M.~Asai, M.~Block, R.~Boll, H.~Brand, D.~Cox, {\em et~al.},
  ``Fusion reaction ca 48+ bk 249 leading to formation of the element ts (z=
  117),'' {\em Physical Review C}, vol.~99, no.~5, p.~054306, 2019.

\bibitem{forsberg2016recoil}
U.~Forsberg, D.~Rudolph, L.-L. Andersson, A.~Di~Nitto, C.~E. D{\"u}llmann,
  C.~Fahlander, J.~Gates, P.~Golubev, K.~Gregorich, C.~Gross, {\em et~al.},
  ``Recoil-$\alpha$-fission and recoil-$\alpha$--$\alpha$-fission events
  observed in the reaction 48ca+ 243am,'' {\em Nuclear Physics A}, vol.~953,
  pp.~117--138, 2016.

\bibitem{oganessian2022first}
Y.~T. Oganessian, V.~Utyonkov, N.~Kovrizhnykh, F.~S. Abdullin, S.~Dmitriev,
  D.~Ibadullayev, M.~Itkis, D.~Kuznetsov, O.~Petrushkin, A.~Podshibiakin, {\em
  et~al.}, ``First experiment at the super heavy element factory: High cross
  section of mc 288 in the am 243+ ca 48 reaction and identification of the new
  isotope lr 264,'' {\em Physical Review C}, vol.~106, no.~3, p.~L031301, 2022.

\bibitem{zhu2024new}
L.~Zhu, ``New model based on coupling the master and langevin equations in the
  study of multinucleon transfer reactions,'' {\em Physics Letters B},
  vol.~849, p.~138423, 2024.

\end{thebibliography}

\end{document}